%Paper: hep-ph/9408333
%From: FARAGGI@sns.ias.edu
%Date: 21 Aug 1994 15:32:00 -0400 (EDT)

% Printing instructions:
%       This paper needs the macro packages phyzzx.tex and tables.tex
%       The tables should be stripped off and printed separately.
%
%
\input phyzzx
\tolerance=1000
\sequentialequations
\def\rl{\rightline}

\def\t1{{\tilde 1}}

\def\AEF{A.E. Faraggi}
\def\DVN{D.V. Nanopoulos}

\def\SUSY{supersymmetry }

\def\NPB#1#2#3{Nucl. Phys. B {\bf#1} (19#2) #3}
\def\PLB#1#2#3{Phys. Lett. B {\bf#1} (19#2) #3}
\def\PRD#1#2#3{Phys. Rev. D {\bf#1} (19#2) #3}

\def\PRT#1#2#3{Phys. Rep. {\bf#1} (19#2) #3}

\REF\SUSY{For reviews see,
H.P. Nilles, \PRT{110}{84}{1}; \DVN~ and Lahanas, \PRT{145}{87}{1}.}
\REF\GSW{M. Green, J. Schwarz and E. Witten, Superstring Theory, 2 vols.,
Cambridge University Press, 1987.}
\REF\DKMNS{M.Dine, V. Kaplunovsky, M. Mangano, C. Nappi and N. Seiberg,
\NPB{259}{85}{549}.}
\REF\SUTHREE{B. Greene {\it{el al.}},
Phys.Lett.{\bf B180} (1986) 69;
Nucl.Phys.{\bf B278} (1986) 667;  {\bf B292} (1987) 606;
R. Arnowitt and  P. Nath, Phys.Rev.{\bf D39} (1989) 2006; {\bf D42}
(1990) 2498; Phys.Rev.Lett. {\bf 62} (1989) 222.}
\REF\FBV{J. Ellis, J. Lopez and D.V. Nanopoulos, \PLB{252}{90}{53},
G. Leontaris and T. Tamvakis, \PLB{260}{91}{333}.}
\REF\NRT{\AEF, \NPB{403}{93}{101}, hep-th/9208023;
IASSNS--94--18, hepph/9403312, Nucl. Phys. {\bf B}, in press.}
\REF\FIQ{A. Font, L.E. Iba{\~n}ez and F. Quevedo,
 Phys.Lett.{\bf B228} (1989) 79.}
\REF\FNY{\AEF, D.V. Nanopoulos and K. Yuan, \NPB{335}{90}{347}.}
\REF\FH{\AEF, \PLB{245}{90}{435};
        \AEF~and E. Halyo, \PLB{307}{93}{311}, hep-th/9303060.}
\REF\FFF{I. Antoniadis, C. Bachas, and C. Kounnas, Nucl.Phys.{\bf B289}
(1987) 87; I. Antoniadis and C. Bachas, Nucl.Phys.{\bf B298} (1988)
586; H. Kawai, D.C. Lewellen, and S.H.-H. Tye,  Nucl.Phys.{\bf B288} (1987) 1;
R. Bluhm, L. Dolan, and P. Goddard, Nucl.Phys.{\bf B309} (1988) 330.}
\REF\REVAMP{I. Antoniadis, J. Ellis, J. Hagelin, and \DVN, \PLB{231}{89}{65};
I. Antoniadis, G. K. Leontaris and J. Rizos, \PLB{245}{90}{161}.}
\REF\SLM{\AEF, \NPB{387}{92}{239}, hep-th/9208024.}
\REF\EU{\AEF,  \PLB{278}{92}{131}.}
\REF\TOP{\AEF, \PLB{274}{92}{47}.}
\REF\PRICE{I. Antoniadis, J. Ellis, S. Kelley, and \DVN, \PLB{272}{91}{31};
J. Lopez, D.V. Nanopoulos and K. Yuan, \NPB{399}{93}{654}, hep-th/9203025.}
\REF\FOC{\AEF, \PLB{326}{94}{62}, hep-ph/9311312.}
\REF\LYKKEN{S. Chaudhuri, S. Chung, and J. D.
Lykken, FERMILAB-PUB-94-137-T, Talk given at 2nd IFT Workshop on Yukawa
Couplings and the Origins of Mass, Gainesville, Feb 1994,
hep-ph/9405374.}
\REF\DSW{M. Dine, N. Seiberg and E. Witten, \NPB{289}{87}{585}.}
\REF\RX{There are numerous papers on this subject.  A partial
list includes L.J. Hall and M. Suzuki, \NPB{231}{84}{419};
I.H. Lee, \NPB{246}{84}120;
S. Dawson, \NPB{261}{85}{297} 297; R. Barbieri and A. Masiero
\NPB{267}{86}{679};
V. Barger, G.F. Giudice, and T. Han, \PRD{40}{89}{2987};
S. Dimopoulos {\it et al.}, \PRD{40}{89}{2987};
H. Dreiner and G.G. Ross, \NPB{365}{91}{597}.}
\REF\FM{\AEF, \NPB{407}{93}{57}, hep-ph/9210256;
	 \AEF~ and E.Halyo, \PLB{307}{93}{305}, hep-ph/9301261;
          		    \NPB{416}{94}{63}, hep-ph/9306235.}
\REF\DOLAN{L. Dolan and S. Horvath , \NPB{416}{94}{87}.}
\REF\DKV{L. Dixon, V. Kaplunovsky and C. vafa, \NPB{294}{87}{43}.}
\REF\DV{N.V. Krasnikov, \PLB{193}{87}{37}; L. Dixon, in Proc. A.P.S.
DPF Meeting, Houston, TX. 1990; T. Taylor, \PLB{252}{90}{59}.}
\REF\BD{T. Banks and L. Dixon, \NPB{307}{88}{93}.}

\singlespace
\rl{IASSNS--HEP--94/64}
\rl{\today}
%\rl{July 1994}
\rl{T}
\normalspace
\smallskip
\titlestyle{\bf{Custodial nonabelian gauge symmetries in realistic
superstring derived models}}
\author{Alon E. Faraggi
{\footnote\dag{Work supported by an SSC fellowship.
               e--mail address: faraggi@sns.ias.edu}}}
\smallskip
\centerline {School of Natural Sciences}
\centerline {Institute for Advanced Study}
\centerline {Olden Lane}
\centerline {Princeton, NJ 08540}

\titlestyle{ABSTRACT}
A well known problem in supersymmetric models is the presence of,
lepton and baryon number violating, dimension four operators.
The traditional R parity solution may not be suitable if
one tries to incorporate supersymmetry into a Planck scale theory.
I propose a different solution in the context of realistic string models.
I show that realistic string models can give rise to custodial
nonabelian gauge symmetries under which only the leptons or quarks transform.
I explain how such symmetries arise in a class of free fermionic models that
are based on $Z_2\times Z_2$ orbifold with standard embedding.
The custodial symmetries forbid proton decay from dimension four operators
while allowing R parity violation.
\centerline{}

\singlespace
\vskip 0.5cm
\nopagenumbers
\pageno=0
\endpage
\normalspace
\pagenumbers

\bigskip
%\centerline{\bf 1. Introduction}

Supersymmetry [\SUSY]
is a phenomenologically appealing extension of the Standard
Model. While LEP precision data strongly constrain other attractive extensions
of the Standard Model, supersymmetry below the TeV scale is in good
agreement with all experiments to date. Moreover, it is derived from
superstring theory and provides a solution to the gauge hierarchy problem.
However, despite this attractive properties, supersymmetry gives rise
to dimension four, baryon and lepton violating, operators that result
in fast proton decay. The dangerous dimension four operators
are forbidden by postulating an extra matter parity symmetry.
In superstring theory [\GSW] the problem is more severe because such a symmetry
cannot be imposed at will. Early in the days of string inspired
phenomenology it was pointed out that matter parity was not
automatic in most string--derived models and that renormalizable
dimension four operators were present in generic string vacua [\DKMNS].
By going to a particular symmetric vacuum, these operators could often be
avoided [\SUTHREE]. However to produce a realistic low energy mass spectrum
it is, in general, necessary to perturb away from the symmetric
points in moduli space. When perturbing away from the symmetric point,
it is difficult to envision how one can control
the absence of dimension four operators.

In $SO(10)$ based models absence of a cubic sixteen operator forbid
the dimension four operators. However, the dangerous dimension four
operators may still be induced from quartic sixteen operators,
if one of the spinorial sixteen of $SO(10)$ gets a GUT or Planck scale VEV,
as such a VEV breaks matter parity.
In terms of standard model multiplets the dangerous operators are
$${\eta_1}({u_{L}^C}{d_{L}^C}{d_{L}^C}N_L^c)\Phi+
{\eta_2}({d_{L}^C}QLN_L^c)\Phi,\eqno(1)$$
where $N_L^c$ is the Standard Model singlet in the $16$ of $SO(10)$.
$\Phi$ is a combination of fields which
fixes the string selection rules and gets a VEV of $O(M/10)$, where
$M=M_{Pl}/2\sqrt{8\pi}$. From Eq. (1), it is seen that the ratio
${\langle{N_L^c}\rangle}/{M}$ controls the rate
of proton decay. A search through nonrenormalizable
terms shows that terms of the form of Eq. (1) are, in general,
generated in string models [\FBV,\NRT].
An additional
$U(1)_{Z^\prime}$ that is unbroken down to low energies
suppresses the dangerous dimension four operators in $SO(10)$ based
models. The  problem with this solution is that often the scale of
$U(1)_{Z^\prime}$ breaking is the scale of a see--saw mechanism that
suppresses the left--handed neutrino masses. While in field theory
this scale can be near the TeV scale, in superstring models,
because the terms in the seesaw mass matrix are usually obtained
from nonrenormalizable terms, the scale of $U(1)_{Z^\prime}$ breaking
is often required to be much higher [\FH].

In this paper I propose that in string theory a different mechanism
is required. In this mechanism custodial nonabelian symmetries are
obtained under which only the leptons or quarks transform. The
custodial symmetry then allows only baryon or lepton violating
dimension four operators but not both. Therefore with the
custodial nonabelian symmetries R--parity may be broken close to the
Planck scale but proton decay from dimension four operators is
suppressed. I construct a toy model to illustrate this mechanism. In this model
the Standard Model leptons transform under a custodial $SU(2)$ symmetry.
In the toy model the custodial symmetries arise due to additional
space--time vector bosons that are obtained from twisted sectors.
These twisted sectors are obtained from boundary condition vectors
that break the $SO(10)$ symmetry and correspond to ``Wilson lines''
in the orbifold formulation. Contrary to the gauged $B-L$ symmetry that
arises in string models solely from the world--sheet gauge degrees of freedom
[\FIQ,\FNY], the custodial symmetries arise from mixture of the gauge
and internal degrees of freedom. I discuss some additional aspects
of similar extended symmetries and their possible
phenomenological implications.

\bigskip
%\centerline{\bf 2. The superstring models }

The superstring models that I present are constructed in the
free fermionic formulation [\FFF]. In this formulation all the degrees
of freedom needed to cancel
the conformal anomaly are represented
in terms of internal free fermions propagating
on the string world--sheet.
Under parallel transport around a noncontractible loop,
the fermionic states pick up a phase.
Specification of the phases for all world--sheet fermions
around all noncontractible loops contributes
to the spin structure of the model.
The possible spin structures are constrained
by string consistency requirements
(e.g. modular invariance).
A model is constructed by choosing a set of boundary condition vectors,
which satisfies
the modular invariance constraints.
The basis vectors, $b_k$, span a finite
additive group $\Xi=\sum_k{{n_k}{b_k}}$
where $n_k=0,\cdots,{{N_{z_k}}-1}$.
The physical massless states in the Hilbert space of a given sector
$\alpha\in{\Xi}$, are obtained by acting on the vacuum with
bosonic and fermionic operators and by
applying the generalized GSO projections. The $U(1)$
charges, Q(f), with respect to the unbroken Cartan generators of the four
dimensional gauge group, which are in one to one correspondence with the $U(1)$
currents ${f^*}f$ for each complex fermion f, are given by:
 $${Q(f) = {1\over 2}\alpha(f) + F(f)}\eqno (2)$$
where $\alpha(f)$ is the boundary condition of the world--sheet fermion $f$
 in the sector $\alpha$, and
$F_\alpha(f)$ is a fermion number operator counting each mode of
$f$ once (and if $f$ is complex, $f^*$ minus once).
For periodic fermions,
$\alpha(f)=1$, the vacuum is a spinor in order to represent the Clifford
algebra of the corresponding zero modes. For each periodic complex fermion $f$
there are two degenerate vacua ${\vert +\rangle},{\vert -\rangle}$ ,
annihilated by the zero modes $f_0$ and
${{f_0}^*}$ and with fermion numbers  $F(f)=0,-1$, respectively.

The realistic models in the free fermionic formulation are generated by
a basis of boundary condition vectors for all world--sheet fermions
[\FNY,\REVAMP--\PRICE].
The basis is constructed in two stages. The first stage consist
of the NAHE set [\SLM], which is a set of five boundary condition basis
vectors, $\{{{\bf 1},S,b_1,b_2,b_3}\}$. The gauge group after the NAHE set
is $SO(10)\times SO(6)^3\times E_8$ with $N=1$ space--time supersymmetry.
The vector $S$ is the supersymmetry generator and the superpartners of
the states from a given sector $\alpha$ are obtained from the sector
$S+\alpha$. The space--time vector bosons that generate the gauge group
arise from the Neveu--Schwarz sector and from the sector $1+b_1+b_2+b_3$.
The Neveu--Schwarz sector produces the generators of
$SO(10)\times SO(6)^3\times SO(16)$. The sector $1+b_1+b_2+b_3$
produces the spinorial 128 of $SO(16)$ and completes the hidden
gauge group to $E_8$.
The vectors $b_1$, $b_2$ and $b_3$ correspond to the three twisted
sectors in the corresponding orbifold formulation and produce
48 spinorial 16 of $SO(10)$, sixteen from each sector $b_1$,
$b_2$ and $b_3$.

The NAHE set divides the 44 right--moving and 20 left--moving real internal
fermions in the following way: ${\bar\psi}^{1,\cdots,5}$ are complex and
produce the observable $SO(10)$ symmetry; ${\bar\phi}^{1,\cdots,8}$ are
complex and produce the hidden $E_8$ gauge group;
$\{{\bar\eta}^1,{\bar y}^{3,\cdots,6}\}$, $\{{\bar\eta}^2,{\bar y}^{1,2}
,{\bar\omega}^{5,6}\}$, $\{{\bar\eta}^3,{\bar\omega}^{1,\cdots,4}\}$
 give rise to the three horizontal $SO(6)$ symmetries. The left--moving
$\{y,\omega\}$ states are divided to, $\{{y}^{3,\cdots,6}\}$, $\{{y}^{1,2}
,{\omega}^{5,6}\}$, $\{{\omega}^{1,\cdots,4}\}$. The left--moving
$\chi^{12},\chi^{34},\chi^{56}$ states carry the supersymmetry charges.
Each sector $b_1$, $b_2$ and $b_3$ carries periodic boundary conditions
under $(\psi^\mu\vert{\bar\psi}^{1,\cdots,5})$ and one of the three groups:
$(\chi_{12},\{y^{3,\cdots,6}\vert{\bar y}^{3,\cdots6}\},{\bar\eta}^1)$,
$(\chi_{34},\{y^{1,2},\omega^{5,6}\vert{\bar y}^{1,2}{\bar\omega}^{5,6}\},
{\bar\eta}^2)$ and $(\chi_{56},\{\omega^{1,\cdots,4}\vert{\bar\omega}^{1,
\cdots4}\},{\bar\eta}^3)$. The division of the internal fermions is a
reflection of the underlying $Z_2\times Z_2$ orbifold
compactification [\FOC]. The set of internal fermions
${\{y,\omega\vert{\bar y},{\bar\omega}\}^{1,\cdots,6}}$
corresponds to the left--right symmetric conformal field theory of the
heterotic string, or to the six dimensional compactified manifold in a
bosonic formulation. This set of left--right symmetric internal fermions
plays a fundamental role  in the determination of the low energy properties
of the realistic free fermionic models.

The second stage of the basis construction consist of adding three
additional basis vectors to the NAHE set. The three additional basis
vectors correspond to ``Wilson lines'' in the orbifold formulation.
Three additional vectors are needed to reduce the number of generations
to three, one from each sector $b_1$, $b_2$ and $b_3$.
The additional basis vectors distinguish between different models
and determine their low energy properties. The allowed boundary
conditions in the additional basis vectors are constrained
by the string consistency constraints, i.e. modular invariance and
world--sheet supersymmetry. The choice of boundary
conditions to the set of internal fermions
${\{y,\omega\vert{\bar y},{\bar\omega}\}^{1,\cdots,6}}$
determines the low energy properties, like the number of generations,
Higgs doublet--triplet splitting and Yukawa couplings.
The low energy phenomenological requirements impose strong
constraints on the possible assignment of boundary conditions to the
set of of internal world--sheet fermions
${\{y,\omega\vert{\bar y},{\bar\omega}\}^{1,\cdots,6}}$ [\SLM].

For some choices of the additional basis vectors that extend the NAHE
set, there exist a combination
$$X=n_\alpha\alpha+n_\beta\beta+n_\gamma\gamma\eqno(3)$$
for which $X_L\cdot X_L=0$ and $X_R\cdot X_R\ne0$. Such a
combination may produce additional space--time vector
bosons, depending on the choice of GSO phases. For example, in the
model of Ref. [\EU] the combination $X=b_1+b_2+b_3+\alpha+\beta+\gamma$ has
$X_L\cdot X_L=0$ and $X_R\cdot X_R=8$. The space--time vector bosons
from this sector are projected out by the choice of GSO phases,
and this vector combination produces only space--time scalar
supermultiplets. On the other hand in the model of Ref. [\TOP]
with the modified GSO phase
$$c\left(\matrix{\gamma\cr1\cr}\right)=+1\rightarrow
c\left(\matrix{\gamma\cr1\cr}\right)=-1$$
additional space--time vector bosons are obtained from the sector
$1+\alpha+2\gamma$. The gauge group after applying the
GSO projections is
$SU(3)_C\times SU(2)_L\times SU(2)_c\times U(1)_{C^\prime}
\times U(1)_L\times U(1)^5\times SU(5)\times SU(3)\times U(1)$.

The gauge group arises as follows:
the NS sector produces the
generators of $SU(3)_C\times SU(2)_L\times U(1)_C\times U(1)_L
\times U(1)_{1,2,3}\times U(1)_{4,5,6}\times SU(3)\times SO(4)\times
U(1)_H\times U(1)_{7,8,9}$ where
$$\eqalignno{U(1)_C&=Tr U(3)_C~\Rightarrow~
Q_C=\sum_{i=1}^3Q({\bar\psi}^i),&(4a)\cr
U(1)_L&= Tr U(2)_L~\Rightarrow~Q_L=\sum_{i=4}^5Q({\bar\psi}^i),&(4b)\cr
U(1)_H&= Tr U(3)_H~\Rightarrow~Q_H=\sum_{i=5}^7Q({\bar\phi}^i).&(4c)\cr}$$
$U(1)_{1,2,3}$, $U(1)_{4,5,6}$ and $U(1)_{5,6,7}$ arise
from the world--sheet currents ${\bar\eta}^i{\bar\eta}^{i^*}$
$(i=1,2,3)$, ${\bar y}^3{\bar y}^6$, ${\bar y}^1{\bar\omega}^5$
${\bar\omega}^2{\bar\omega}^4$, and ${\bar\phi}^1{\bar\phi}^{1^*}$,
${\bar\phi}^2{\bar\phi}^{2^*}$, ${\bar\phi}^8{\bar\phi}^{8^*}$,
respectively.
The sector $1+b_1+b_2+b_3$ produces the representations
$(3,2)_{-5}\oplus({\bar3},2)_5$ and $2_{-3}\oplus2_3$ of
$SU(3)\times SU(2)_r\times U(1)_{h5}$ and $SU(2)_\ell\times U(1)_{h3}$
respectively, where
$SU(2)_r\times SU(2)_\ell$ are the two $SU(2)$'s in the isomorphism
$SO(4)\sim SU(2)_r\times SU(2)_\ell$.
Thus, the $E_8$ symmetry
reduces to $SU(5)\times SU(3)\times U(1)^2$.
The $U(1)$'s in $SU(5)$ and $SU(3)$ are given by
$U(1)_{h5}=-3U_7+3U_8+U_H-3U_9$ and $U(1)_{h3}=U_7+U_8+U_H+U_9$
respectively.
The remaining $U(1)$ symmetries in the
hidden sector, $U(1)_{7^\prime}$ and $U(1)_{8^\prime}$,
correspond to the world--sheet currents
${\bar\phi}^1{\bar\phi}^{1^*}-{\bar\phi}^8{\bar\phi}^{8^*}$ and
$-2{\bar\phi}^j{\bar\phi}^{j^*}+{\bar\phi}^1{\bar\phi}^{1^*}
+4{\bar\phi}^2{\bar\phi}^{2^*}+{\bar\phi}^8{\bar\phi}^{8^*}$ respectively,
where summation on $j=5,\cdots,7$ is implied.

The sector $1+\alpha+2\gamma$ produces two additional space--time
vector bosons, which are singlets of the nonabelian group but
carry $U(1)$ charges. One combination of the $U(1)$ symmetries
		$$U_C+U_4+U_5+U_6+U_{7^\prime}\eqno(5)$$
is the $U(1)$ of the custodial
$SU(2)$ symmetry. The two space--time vector bosons from the sector
$1+\alpha+2\gamma$ produce the two additional vector bosons of the
custodial $SU(2)$ gauge group. The remaining orthogonal combinations are
$$\eqalignno{U_{C^\prime}&=
{1\over3}U_C-{1\over2}U_{7^\prime},&(6a)\cr
U_{4^\prime}&=U_4-U_6,&(6b)\cr
U_{5^\prime}&=U_4+U_5-2U_6,&(6c)\cr
U_{7^{\prime\prime}}&=U_C-{5\over3}(U_4+U_5+U_6)+U_{7^\prime}.&(6d)\cr}$$
The full massless spectrum now transforms under the final gauge group,
$SU(3)_C\times SU(2)_L\times SU(2)_c\times U(1)_{C^\prime}\times U(1)_L\times
U(1)_{1,2,3}\times U(1)_{4^\prime}\times U(1)_{5^\prime}\times
U(1)_{7^{\prime\prime}}\times U(1)_8$.

(a) The Neveu-Schwarz $O$ sector gives, in addition to  the graviton,
dilaton, antisymmetric tensor and spin 1 gauge bosons,  the
following scalar representations:
$$\eqalignno{
{h_1}&\equiv{[(1,0);(2,-1)]}_{1,0,0,0,0,0}
{\hskip .5cm}
\Phi_{23}\equiv{[(1,0);(1,0)]}_{0,1,-1,0,0,0}&(7a,b)\cr
{h_2}&\equiv{[(1,0);(2,-1)]}_{0,1,0,0,0,0} {\hskip .5cm}
\Phi_{13}\equiv{[(1,0);(1,0)]}_{1,0,-1,0,0,0}&(7c,d)\cr
{h_3}&\equiv{[(1,0);(2,-1)]}_{0,0,1,0,0,0}{\hskip .5cm}
\Phi_{12}\equiv{[(1,0);(1,0)]}_{1,-1,0,0,0,0}
&(7e,f)\cr}$$
(and their conjugates ${\bar h}_1$ etc.).
Finally, the Neveu--Schwarz sector gives rise to three singlet
states that are neutral under all the U(1) symmetries.
$\xi_{1,2,3}:{\hskip .2cm}{\chi^{12}_{1\over2}{\bar\omega}^3_{1\over2}
{\bar\omega}^6_{1\over2}{\vert 0\rangle}_0},$
 ${\chi^{34}_{{1\over2}}{\bar y}_{1\over2}^5{\bar\omega}_{1\over2}^1
{\vert 0\rangle}_0},$
 $\chi^{56}_{1\over2}{\bar y}_{1\over2}^2{\bar y}_{1\over2}^4
{\vert 0\rangle}_0.$

(b) The ${S+b_1+b_2+\alpha+\beta}$ sector gives
$$\eqalignno{h_{45}&\equiv{[(1,0);(2,-1)]}_
{{1\over2},{1\over2},0,0,0,0} {\hskip .5cm}
{h}_{45}^\prime\equiv{[(1,0);(2,-1)]}_
{-{1\over2},-{1\over2},0,0,0,0}&(8a,b)\cr
\Phi_{45}&\equiv{[(1,0);(1,0)]}_
{-{1\over2},-{1\over2},-1,0,0,0}  {\hskip .5cm}
\Phi^{\prime}_{45}\equiv{[(1,0);(1,0)]}_
{-{1\over2},-{1\over2},1,0,0,0}&(8c,d)\cr
\Phi_1&\equiv{[(1,0);(1,0)]}_
{-{1\over2},{1\over2},0,0,0,0} {\hskip .5cm}
\Phi_2\equiv{[(1,0);(1,0)]}_
{-{1\over2},{1\over2},0,0,0,0}&(8e,f)\cr}$$
(and their conjugates ${\bar h}_{45}$, etc.).
The states are obtained by acting on the vacuum
with the fermionic oscillators
${\bar\psi}^{4,5},{\bar\eta}^3,{\bar y}_5,{\bar\omega}_6$,
respectively  (and their complex conjugates for ${\bar h}_{45}$, etc.).

The sectors $b_j\oplus 1+\alpha+2\gamma$
produce the three light generations, one for each of the sectors
$b_j$ $(j=1,2,3)$. The states from these sectors and their decomposition
under the entire gauge group are shown in table 1. From table 1
we see that only the lepton supermultiplets, $\{L, e_L^c, N_L^c\}$
transform as doublets under the custodial $SU(2)$ gauge group
while the quarks are singlets. The remaining
matter states in the massless spectrum and their quantum numbers are
given in table 2.

The model contains three anomalous $U(1)$ symmetries:
Tr${U_1}=24$, Tr${U_2}=24$, Tr${U_3}=24$.
Of the three anomalous $U(1)$s,  two can be rotated by
an orthogonal transformation. One combination remains anomalous and is
uniquely given by: ${U_A}=k\sum_j [{Tr {U(1)_j}}]U(1)_j$,
where $j$ runs over all the
anomalous $U(1)$s.
For convenience, I take $k={1/{24}}$. Therefore,
the anomalous combination
is given by:
$$U_A=U_1+U_2+U_3,{\hskip 3cm}TrQ_A=72.\eqno(9a)$$
The two orthogonal combinations are not unique. Different
choices are related by orthogonal transformations. One choice is given by:
$$\eqalignno{{U^\prime}_1&=U_1-U_2{\hskip .5cm},{\hskip .5cm}
{U^\prime}_2=U_1+U_2-2U_3.&(9b,c)\cr}$$
Together with the other anomaly free $U(1)$s,
they are free from  gauge and gravitational
anomalies. The cancelation of all mixed  anomalies among the $U(1)$s
is a non trivial consistency check of the
massless spectrum of the model.
The ``anomalous" $U(1)_A$ is broken
by the Dine-Seiberg-Witten mechanism [\DSW]
in which some states in the massless spectrum obtain nonvanishing VEVs
that cancel the anomalous $U(1)$ D--term equation.
A particular example, in the
model under consideration, is given by the set
$\{\Phi_{45},\Phi^\prime_{45}\}$ with
$\vert\langle\Phi_{45}\rangle\vert^2=
3\vert\langle\Phi^\prime_{45}\rangle\vert^2=
{{3g^2}\over{16\pi^2}}$.

Nonvanishing VEVs of the states form the sectors $b_j+2\gamma$
break the $U(1)$ symmetries to $U(1)_C\times U(1)_L$. The VEV
of $N_L^c$ breaks the custodial $SU(2)_c$ symmetry and
the remaining $U(1)$ symmetry. The surviving
combination $1/3U(1)_C+1/2U(1)_L$ is the Standard Model weak hypercharge.
Only the leptons $L_j$, $e_j$ and
$N_j$ transform as doublets under the custodial $SU(2)_c$ gauge group
whereas the quarks are singlets of $SU(2)_c$ (see table 1).
Consequently, the term $QLd_L^cN_L^c\Phi$ is allowed while the term
$${u_{L}^C}{d_{L}^C}{d_{L}^C}N_L^c\Phi\eqno(10)$$
is forbidden due to invariance
under $SU(2)_c$.
Thus, the VEV of $N_L^c$ can be of order $M_{Pl}$, and
although it breaks matter parity, it does not imply proton decay
from dimension four operators. While the lepton number violating dimension
four operator, $QLd^c$,
is allowed and may be unsuppressed, baryon number violating
dimension four operators are forbidden. An important implication of R
parity violation is that the lightest supersymmetric particle is
unstable. Analysis of models that
allow this type of matter parity breaking has been extensive and I refer
the interested reader to the literature [\RX]. The Yukawa couplings
$Qd^ch$, $Qu^c{\bar h}$ and $Le^ch$ are invariant under the
custodial $SU(2)_c$ symmetry. Therefore, the same fermion mass textures
are expected to arise as in the models in which the custodial
$SU(2)$ is absent [\NRT,\FM].

A similar mechanism may be possible in the case of superstring flipped $SU(5)$
models [\REVAMP,\PRICE] and other string GUT models [\LYKKEN]. For example,
additional gauge symmetries from twisted sectors
were shown to arise in the flipped $SU(5)$
model of Ref. [\PRICE]. The extended
symmetries arise because of the existence of a sector in the additive group,
of the form of Eq. (3), with $X_L\cdot X_L=0$. Such a sector exist in
the additive group because of the assignment of boundary conditions
to the set of internal world--sheet fermions
$\{y,\omega\vert{\bar y},{\bar\omega}\}^{1,\cdots,6}$. In terms of flipped
$SU(5)$ representations the dimension four operators arise from the
operator $$FF{\bar f}H\eqno(11)$$
where $F$ and $H$ are in the 10 representation of
$SU(5)$ and ${\bar f}$ is in the ${\bar 5}$ representation of $SU(5)$.
The decomposition under standard model representations is:
$F=(Q,d_L^c,N_L^c)$, ${\bar f}=(u_L^c,L)$. The neutral state in $H$
obtains a GUT scale VEV and breaks the $SU(5)\times U(1)$ symmetry
to $SU(3)_C\times SU(2)_L\times U(1)_Y$. However, such a VEV,
in general, generates also the dangerous dimension four operators.
In the presence of custodial nonabelian gauge symmetries, similar
to the one shown to arise in some standard--like models,
the dangerous operators may be forbidden due to the custodial symmetry.

Extended symmetries from twisted sectors may have additional
phenomenological implications. In Ref. [\DOLAN] extended gauge
symmetries from twisted sectors were sought in type II superstring
in order to circumvent the no go theorem of Ref. [\DKV].
Examining the GSO phases in the superstring standard--like models
it is observed that different choices of GSO phases result in different
extensions of the gauge group. For example,
in the model of Ref. [\EU] extended gauge symmetries may arise
from the sector $b_1+b_2+b_3+\alpha+\beta+\gamma+(I)$, where
$I=1+b_1+b_2+b_3$.
With the choice of GSO phases in Ref. [\EU] all the extra gauge
bosons are projected out by the GSO projections. However,
with the modified GSO phases
$$c\left(\matrix{1\cr\gamma\cr}\right)\rightarrow
-c\left(\matrix{1\cr\gamma\cr}\right),
 c\left(\matrix{\alpha\cr\beta\cr}\right)\rightarrow
-c\left(\matrix{\alpha\cr\beta\cr}\right) ~\hbox{and}~
 c\left(\matrix{\gamma\cr\beta\cr}\right)\rightarrow
-c\left(\matrix{\gamma\cr\beta\cr}\right),$$ additional space--time
vector bosons are obtained from the sector
$b_1+b_2+b_3+\alpha+\beta+\gamma+(I)$ .
The sector $b_1+b_2+b_3+\alpha+\beta+\gamma+(I)$
produces the representations $3_1+3_{-1}$ of $SU(3)_H$, where
one of the $U(1)$ combinations is the $U(1)$ in the decomposition
of $SU(4)$ under $SU(3)\times U(1)$.
In this case the hidden $SU(3)_H$ gauge group is extended to
$SU(4)_H$. Thus, the hidden sector contains two nonabelian factors
$SU(5)\times SU(4)$. The possibility of extending the hidden sector
gauge group from twisted sectors may be instrumental in trying to
implement the dilaton stabilization mechanism of Ref. [\DV]. It should be
noted that in the model of Ref. [\TOP] the additional space--time vector
bosons from the sector $1+\alpha+2\gamma$ can be projected from the massless
spectrum as well. However, if we require $N=1$ space--time supersymmetry
then the projection of the additional gauge bosons is correlated
with the presence of Higgs doublets from the sector
$b_1+b_2+\alpha+\beta$ in the massless spectrum. This is easily
seen by substituting the vectors $\alpha$ and $\beta$ in the basis
with the vectors $b_1+b_2+\alpha+\beta$ and $1+\alpha+2\gamma$.
It is observed that the intersection between these two vectors
is empty. Therefore, the GSO projections correlate between
Higgs doublets with extra gauge bosons or Higgs triplets without
extra gauge bosons. The only other alternative that was found
to project out the extra gauge bosons is by projecting out the last surviving
gravitino, thus breaking supersymmetry at the Planck scale. This is
achieved by modifying the phase
$c\left(\matrix{S\cr\alpha\cr}\right)\rightarrow
-c\left(\matrix{S\cr\alpha\cr}\right)$. It would be of
interest to examine whether this class of nonsupersymmetric models can
produce realistic phenomenology and vanishing cosmological constant.
Finally, the extended gauge symmetries provide some freedom in the
definition of the weak hypercharge. We may still define the
weak hypercharge to be $U(1)_Y=1/3U(1)_C+1/2U(1)_L$ with the
standard $SO(10)$ embedding. However, we may also define it
as $U(1)_Y=U(1)_{C^\prime}+1/2U(1)_L$. With these two definitions,
the weak hypercharge of the quark and leptons are the same.
This freedom may be instrumental in trying to understand the
disparity between the gauge coupling unification scale in the
Minimal Supersymmetric Standard Model (MSSM) and the string unification scale.

\bigskip
%\centerline{\bf 2. Conclusions}

In this paper I have shown how additional gauge bosons may appear in
realistic superstring derived models from twisted sectors.
In the free fermionic models that are based on $Z_2\times Z_2$
orbifold with standard embedding the additional gauge bosons
may give rise to custodial nonabelian gauge symmetries,
under which only the leptons transform. In the fermionic models
the extended symmetries arise because of the asymmetry of the boundary
conditions in the vectors $\alpha$, $\beta$, $\gamma$ between the left,
$\{y,\omega\}$, and right $\{{\bar y},{\bar\omega}\}$, internal world--sheet
fermions. In the orbifold formulation the extended symmetries
should be regarded as arising due to the asymmetry
of the twists $\alpha$, $\beta$, $\gamma$. As a result of the custodial
symmetry, dimension four baryon number violating operators are forbidden
to all orders of nonrenormalizable terms while dimension four lepton number
violating operators are allowed. Consequently, R--parity may be broken close
to the Planck scale, but proton decay cannot be mediated by dimension four
operators. It should be noted that the custodial symmetry also forbid the
dimension five operator $QQQL$. Combined with the selection imposed by the
left--moving global $U(1)$ symmetries, this implies that dimension five
operators that may result in proton decay are forbidden to all orders
of nonrenormalizable terms.
The effective low energy superpotential may contain the dimension
four lepton number violating operator, $QLd^c$, while the dimension
four baryon violating operator is forbidden. Due to the
absence of continuous global symmetries in superstring theory [\BD] and
possibly in any Planck scale theory, a mechanism similar in nature
to the one proposed in this paper, may be the only possible
avenue to avoid proton decay in supersymmetric Planck scale theories.
The possible R parity violating terms are specified explicitly in
specific models. Consequently, the specific low energy predictions
are expected to be different from the low energy phenomenology of the MSSM.

\refout
\vfill
\eject
\input tables.tex
\nopagenumbers
\magnification=1000
\baselineskip=18pt
%\special{landscape}
%\hoffset=1.25truein
%\nopagenumbers
\magnification=1000
%\font\normalroman=cmr10
%\font\style=cmr7
%\style
\tolerance=1200

%\fontdimen12\fivesy=0pt

%\textfont0=\sevenrm
%\scriptfont0=\fiverm
%\textfont1=\seveni
%\scriptfont1=\fivei
%\textfont2=\sevensy
%\scriptfont2=\fivesy
{\hfill
{\begintable
\  \ \|\ ${\psi^\mu}$ \ \|\ $\{{\chi^{12};\chi^{34};\chi^{56}}\}$  \ \|\
{${\bar\psi}^1$, ${\bar\psi}^2$, ${\bar\psi}^3$,
${\bar\psi}^4$, ${\bar\psi}^5$, ${\bar\eta}^1$,
${\bar\eta}^2$, ${\bar\eta}^3$} \ \|\
{${\bar\phi}^1$, ${\bar\phi}^2$, ${\bar\phi}^3$, ${\bar\phi}^4$,
${\bar\phi}^5$, ${\bar\phi}^6$, ${\bar\phi}^7$, ${\bar\phi}^8$} \crthick
${\alpha}$
\|\ 0 \|
$\{0,~0,~0\}$  \|
1, ~~1, ~~1, ~~0, ~~0, ~~0, ~~0, ~~0 \|
1, ~~1, ~~1, ~~1, ~~0, ~~0, ~~0, ~~0 \nr
${\beta}$
\|\ 0 \| $\{0,~0,~0\}$  \|
1, ~~1, ~~1, ~~0, ~~0, ~~0, ~~0, ~~0 \|
1, ~~1, ~~1, ~~1, ~~0, ~~0, ~~0, ~~0 \nr
${\gamma}$
\|\ 0 \|
$\{0,~0,~0\}$  \|
 ~~$1\over2$, ~~$1\over2$, ~~$1\over2$, ~~$1\over2$,
{}~~$1\over2$, ~~$1\over2$, ~~$1\over2$, ~~$1\over2$ \|
$1\over2$, ~~0, ~~1, ~~1,
{}~~$1\over2$,
{}~~$1\over2$, ~~$1\over2$, ~~0 \endtable}
\hfill}
\smallskip
{\hfill
{\begintable
\  \ \|\
${y^3y^6}$,  ${y^4{\bar y}^4}$, ${y^5{\bar y}^5}$,
${{\bar y}^3{\bar y}^6}$
\ \|\ ${y^1\omega^6}$,  ${y^2{\bar y}^2}$,
${\omega^5{\bar\omega}^5}$,
${{\bar y}^1{\bar\omega}^6}$
\ \|\ ${\omega^1{\omega}^3}$,  ${\omega^2{\bar\omega}^2}$,
${\omega^4{\bar\omega}^4}$,  ${{\bar\omega}^1{\bar\omega}^3}$ \crthick
${\alpha}$ \|
1, ~~~1, ~~~~1, ~~~~0 \|
1, ~~~1, ~~~~1, ~~~~0 \|
1, ~~~1, ~~~~1, ~~~~0 \nr
${\beta}$ \|
0, ~~~1, ~~~~0, ~~~~1 \|
0, ~~~1, ~~~~0, ~~~~1 \|
1, ~~~0, ~~~~0, ~~~~0 \nr
${\gamma}$ \|
0, ~~~0, ~~~~1, ~~~~1 \|\
1, ~~~0, ~~~~0, ~~~~0 \|
0, ~~~1, ~~~~0, ~~~~1 \endtable}
\hfill}
\smallskip
\parindent=0pt
\hangindent=39pt\hangafter=1
%\normalroman
\baselineskip=18pt
{{\it Table 1.} A three generations standard--like model that produces
 a custodial $SU(2)$ symmetry.}

\vskip 2cm

\vfill
\eject

\hbox
{\hfill
{\begintable
\ F \ \|\ SEC \ \|\ $SU(3)_C$ $\times$ $SU(2)_L$ $\times$ $SU(2)_c$ \ \|\
$Q_{C^\prime}$ & $Q_L$ & $Q_1$ & $Q_2$ & $Q_3$ & $Q_{4^\prime}$ &
$Q_{5^\prime}$
\ \|\ $SU(5)$ $\times$ $SU(3)$ \ \|\ $Q_{7^{\prime\prime}}$ & $Q_8$  \crthick
$L_1$ \|\
$b_1$ $\oplus$
\|(1,2,2)\|
$-{1\over2}$ & ~~0 & ~~${1\over 2}$ & ~~0 & ~~0 & $-{1\over 2}$ &  $-{1\over2}$
\|(1,1)\| $-{2\over3}$ & ~~0   \nr
$Q_1$  \|\
${1+\alpha+2\gamma}$
\|(3,2,1)\|
{}~~${1\over6}$ & ~~0 & ~~${1\over 2}$ & ~~0 & ~~0 & $-{1\over 2}$ &
$-{1\over2}$
\|(1,1)\| ~~${4\over3}$ & ~~0  \nr
$d_1$ \|\
\|(${\bar 3}$,1,1)\|
$-{1\over6}$ & ~~1 & ~~${1\over 2}$ & ~~0  & ~~0 & ~~${1\over 2}$ &
{}~~${1\over2}$
\|(1,1)\| $-{4\over 3}$ & ~~0   \nr
$N_1$ \|\
\|(1,1,2)\|
{}~~${1\over2}$ & $-1$ & ~~${1\over 2}$ & ~~0 & ~~0 & ~~${1\over 2}$ &
{}~~${1\over2}$
\|(1,1)\| ~~${2\over3}$ & ~~0 \nr
$e_1$ \|\
\|(1,1,2)\|
{}~~${1\over2}$ & ~~1 & ~~${1\over 2}$ & ~~0 & ~~0 & ~~${1\over 2}$ &
{}~~${1\over2}$
\|(1,1)\| ~~${2\over3}$ & ~~0 \nr
$u_1$ \|\
\|(${\bar 3}$,1,1)\|
$-{1\over6}$ & $-1$ & ~~${1\over 2}$ & ~~0 & ~~0 & ~~${1\over 2}$ &
{}~~${1\over2}$
\|(1,1)\| $-{4\over 3}$ & ~~0  \crthick
$L_2$ \|\
$b_2$ $\oplus$
\|(1,2,2)\|
$-{1\over2}$ & ~~0 & ~~0 & ~~${1\over 2}$ & ~~0 & ~~${1\over 2}$ &
$-{1\over2}$
\|(1,1)\| $-{2\over3}$ & ~~0   \nr
$Q_2$  \|\
${1+\alpha+2\gamma}$
\|(3,2,1)\|
{}~~${1\over6}$ & ~~0 & ~~0 & ~~${1\over 2}$ & ~~0 & ~~${1\over 2}$ &
$-{1\over2}$
\|(1,1)\| ~~${4\over3}$ & ~~0  \nr
$d_2$ \|\
\|(${\bar 3}$,1,1)\|
$-{1\over6}$ & ~~1 & ~~0 & ~~${1\over 2}$ & ~~0 & $-{1\over 2}$ &
{}~~${1\over2}$
\|(1,1)\| $-{4\over 3}$ & ~~0   \nr
$N_2$ \|\
\|(1,1,2)\|
{}~~${1\over2}$ & $-1$ & ~~0 & ~~${1\over 2}$ & ~~0 & $-{1\over 2}$ &
{}~~${1\over2}$
\|(1,1)\| ~~${2\over3}$ & ~~0 \nr
$e_2$ \|\
\|(1,1,2)\|
{}~~${1\over2}$ & ~~1 & ~~0 & ~~${1\over 2}$ & ~~0 & $-{1\over 2}$ &
{}~~${1\over2}$
\|(1,1)\| ~~${2\over3}$ & ~~0 \nr
$u_2$ \|\
\|(${\bar 3}$,1,1)\|
$-{1\over6}$ & $-1$ & ~~0 & ~~${1\over 2}$ & ~~0 & $-{1\over 2}$ &
{}~~${1\over2}$
\|(1,1)\| $-{4\over 3}$ & ~~0   \crthick
$L_3$ \|\
${b_3}$ $\oplus$ \|(1,2,2)\|
$-{1\over2}$ & ~~0 & ~~0 & ~~0 &  ~~${1\over 2}$ & ~~0 & ~~1
\|(1,1)\| $-{2\over3}$ & ~~0 \nr
$Q_3$ \|\
${1+\alpha+2\gamma}$
\|(3,2,1)\|
{}~~${1\over6}$ & ~~0 & ~~0 & ~~0 &  ~~${1\over2}$ & ~~0  & ~~1
\|(1,1)\| ~~${4\over 3}$ & ~~0  \nr
$d_3$ \|\
\|(${\bar 3}$,1,1)\|
$-{1\over6}$ & ~~1 & ~~0 & ~~0 & ~~${1\over 2}$ & ~~0 & ${-1}$
\|(1,1)\| $-{4\over 3}$ & ~~0   \nr
$N_3$ \|\
\|(1,1,2)\|
{}~~${1\over2}$ & $-1$ & ~~0 & ~~0 & ~~${1\over 2}$ & ~~0 & ${-1}$
\|(1,1)\| ~~${2\over 3}$ & ~~0  \nr
$e_3$ \|\
\|(1,1,2)\|
{}~~$1\over2$ & ~~1 & ~~0 & ~~0 &  ~~${1\over 2}$ & ~~0 & ${-1}$
\|(1,1)\| ~~${2\over3}$ & ~~0   \nr
${u}_3$ \|\
\|(${\bar 3}$,1,1)\|
$-{1\over6}$ & $-1$ & ~~0 & ~~0 & ~~${1\over 2}$ & ~~0 & ${-1}$
\|(1,1)\| $-{4\over3}$ & ~~0
\endtable}
\hfill}
\bigskip
\parindent=0pt
\hangindent=39pt\hangafter=1
{\it Table 2.}
Massless states and their quantum numbers in the model of table 1.

\vfill
\eject

\input tables.tex
\nopagenumbers
\magnification=1000
\baselineskip=18pt
\hbox
{\hfill
{\begintable
\ F \ \|\ SEC \ \|\ $SU(3)_C$ $\times$ $SU(2)_L$ $\times$ $SU(2)_c$ \ \|\
$Q_{C^\prime}$ & $Q_L$ & $Q_1$ & $Q_2$ & $Q_3$ & $Q_{4^\prime}$ &
$Q_{5^\prime}$
\ \|\ $SU(5)$ $\times$ $SU(3)$ \ \|\ $Q_{7^{\prime\prime}}$ & $Q_8$  \crthick
$V_1$ \|\ $b_1+2\gamma$
\|(1,1,1)\|
{}~~${1\over4}$ & ~~0 & ~~0 & ~~${1\over 2}$ & ~~${1\over 2}$
& ~~${1\over 2}$ & ~~${1\over 2}$
\|(1,3)\| $-{4\over 3}$ & ~~${5\over2}$   \nr
${\bar V}_1$ \|\
\|(1,1,1)\|
$-{1\over4}$ & ~~0 & ~~0 & ~~${1\over 2}$ & ~~${1\over 2}$
& $-{1\over 2}$ & $-{1\over 2}$
\|(1,${\bar3}$)\| ~~${4\over3}$ & $-{5\over2}$ \nr
${T}_1$ \|\
\|(1,1,1)\|
{}~~${1\over4}$ & ~~0 & ~~0 & ~~${1\over 2}$ & ~~${1\over 2}$
& ~~${1\over 2}$ & ~~${1\over 2}$
\|(5,1)\| $-{4\over 3}$ & $-{3\over2}$ \nr
${\bar T}_1$ \|\
\|(1,1,1)\|
$-{1\over4}$ & ~~0 & ~~0 & ~~${1\over 2}$ & ~~${1\over 2}$
& $-{1\over 2}$ & $-{1\over 2}$
\|(${\bar5}$,1)\| ~~${4\over3}$ & ~~${3\over2}$ \cr
$V_2$ \|\ $b_2+2\gamma$
\|(1,1,1)\|
{}~~${1\over4}$ & ~~0 & ~~${1\over2}$ & ~~0 & ~~${1\over 2}$ & $-{1\over2}$
& ~~${1\over2}$
\|(1,3)\| $-{4\over 3}$ & ~~${5\over2}$   \nr
${\bar V}_2$ \|\
\|(1,1,1)\|
$-{1\over4}$ & ~~0 & ~~${1\over 2}$ & ~~0 & ~~${1\over 2}$ & ~~${1\over2}$
& $-{1\over2}$
\|(1,${\bar 3}$)\| ~~${4\over3}$ & $-{5\over2}$ \nr
$T_2$ \|\
\|(1,1,1)\|
{}~~${1\over4}$ & ~~0 & ~~${1\over2}$ & ~~0 & ~~${1\over 2}$ & $-{1\over2}$
& ~~${1\over2}$
\|(5,1)\| $-{4\over 3}$ & $-{3\over2}$ \nr
${\bar T}_2$ \|\
\|(1,1,1)\|
$-{1\over4}$ & ~~0 & ~~${1\over 2}$ & ~~0 & ~~${1\over 2}$ & ~~${1\over2}$
& $-{1\over2}$
\|(${\bar5}$,1)\| ~~${4\over3}$ & ~~${3\over2}$ \cr
$V_3$ \|\ $b_3+2\gamma$
\|(1,1,1)\|
{}~~${1\over4}$ & ~~0 & ~~${1\over 2}$ & ~~${1\over 2}$ & ~~0 & ~~0 & $-1$
\|(1,3)\| $-{4\over 3}$ & ~~${5\over2}$   \nr
${\bar V}_3$ \|\
\|(1,1,1)\|
$-{1\over4}$ & ~~0 & ~~${1\over 2}$ & ~~${1\over 2}$ & ~~0 & ~~0 & ~~1
\|(1,${\bar 3}$)\| ~~${4\over3}$ & $-{5\over2}$ \nr
${T}_3$ \|\
\|(1,1,1)\|
{}~~${1\over4}$ & ~~0 & ~~${1\over 2}$ & ~~${1\over 2}$ & ~~0 & ~~0 & $-1$
\|(5,1)\| $-{4\over 3}$ & $-{3\over2}$ \nr
${\bar T}_3$ \|\
\|(1,1,1)\|
$-{1\over4}$ & ~~0 & ~~${1\over 2}$ & ~~${1\over 2}$ & ~~0 & ~~0 & ~~1
\|(${\bar 5}$,1)\| ~~${4\over3}$ & ~~${3\over2}$ \cr
$D_1$ \|\ $b_1+b_3+\beta\pm\gamma$
\|(3,1,1)\|
{}~~${5\over{24}}$ & ~~${1\over 2}$ & $-{1\over4}$ & ~~${1\over4}$ &
$-{1\over4}$ & ~~0 & ~~0
\|(1,1)\| ~~0 & $-{{15}\over4}$   \nr
${\bar D}_1$ \|\
\|(${\bar 3}$,1,1)\|
$-{5\over{24}}$ & $-{1\over2}$ & ~~${1\over4}$ & $-{1\over4}$ & ~~${1\over4}$
& ~~0 & ~~0
\|(1,1)\| ~~0 & ~~${{15}\over4}$ \nr
$H_1$ \|\
\|(1,1,1)\|
$-{5\over8}$ & ~~${1\over 2}$ & $-{1\over4}$ & ~~${1\over4}$ &
$-{1\over4}$ & ~~0 & ~~0
\|(1,${\bar 3}$)\| ~~0 & ~~${{5}\over4}$ \nr
${\bar H}_1$ \|\
\|(1,1,1)\|
{}~~${5\over8}$ & $-{1\over 2}$ & ~~${1\over4}$ & $-{1\over4}$ &
{}~~${1\over4}$  & ~~0 & ~~0
\|(1,3)\| ~~0 & $-{5\over4}$ \cr
$D_2$ \|\ $b_2+b_3+\beta\pm\gamma$
\|(3,1,1)\|
{}~~${5\over{24}}$ & ~~${1\over 2}$ & ~~${1\over4}$ & $-{1\over4}$ &
$-{1\over4}$ & ~~0 & ~~0
\|(1,1)\| ~~0 & $-{{15}\over4}$   \nr
${\bar D}_2$ \|\
\|(${\bar 3}$,1,1)\|
$-{5\over{24}}$ & $-{1\over2}$ & $-{1\over4}$ & ~~${1\over4}$ & ~~${1\over4}$
& ~~0 & ~~0
\|(1,1)\| ~~0 & ~~${{15}\over4}$ \nr
$H_2$ \|\
\|(1,1,1)\|
$-{5\over8}$ & ~~${1\over 2}$ & ~~${1\over4}$ & $-{1\over4}$ &
$-{1\over4}$ & ~~0 & ~~0
\|(1,${\bar 3}$)\| ~~0 & ~~${{5}\over4}$ \nr
${\bar H}_2$ \|\
\|(1,1,1)\|
{}~~${5\over8}$ & $-{1\over 2}$ & $-{1\over4}$ & ~~${1\over4}$ &
{}~~${1\over4}$  & ~~0 & ~~0
\|(1,3)\| ~~0 & $-{5\over4}$ \cr
$\ell_1$ \|\ $1+b_1+\alpha+2\gamma$
\|(1,2,1)\|
{}~~${1\over2}$ & ~~0 & $-{1\over 2}$ & ~~0 & ~~0 & $-{1\over2}$
& $-{1\over2}$
\|(1,1)\| $-{8\over3}$ & ~~0  \nr
$S_1$ \|\
\|(1,1,1)\|
$-{1\over{2}}$ & $-1$ & $-{1\over2}$ & ~~0 & ~~0 & ~~${1\over2}$ &
{}~~${1\over2}$
\|(1,1)\| ~~${8\over3}$ & ~~0 \nr
$S^\prime_1$ \|\
\|(1,1,1)\|
$-{1\over2}$ & ~~1 & $-{1\over2}$ & ~~0 & ~~0 & ~~${1\over2}$ & ~~${1\over2}$
\|(1,1)\| ~~${8\over3}$ & ~~0 \cr
$\ell_2$ \|\ $1+b_2+\alpha+2\gamma$
\|(1,2,1)\|
{}~~${1\over2}$ & ~~0 & ~~0 & $-{1\over 2}$ & ~~0 & ~~${1\over2}$ &
$-{1\over2}$
\|(1,1)\| $-{8\over3}$ & ~~0  \nr
$S_2$ \|\
\|(1,1,1)\|
$-{1\over{2}}$ & $-1$ & ~~0 & $-{1\over2}$ & ~~0 & $-{1\over2}$ & ~~${1\over2}$
\|(1,1)\| ~~${8\over3}$ & ~~0 \nr
$S^\prime_2$ \|\
\|(1,1,1)\|
$-{1\over2}$ & ~~1 & ~~0 & $-{1\over2}$ & ~~0 & $-{1\over2}$ & ~~${1\over2}$
\|(1,1)\| ~~${8\over3}$ & ~~0 \cr
$\ell_3$ \|\ $1+b_3+\alpha+2\gamma$
\|(1,2,1)\|
{}~~${1\over2}$ & ~~0 & ~~0 & ~~0 & $-{1\over 2}$ & ~~0 & ~~1
\|(1,1)\| $-{8\over3}$ & ~~0  \nr
$S_3$ \|\
\|(1,1,1)\|
$-{1\over{2}}$ & $-1$ & ~~0 & ~~0 & $-{1\over2}$ & ~~0 & $-1$
\|(1,1)\| ~~${8\over3}$ & ~~0 \nr
$S^\prime_3$ \|\
\|(1,1,1)\|
$-{1\over2}$ & ~~1 & ~~0 & ~~0 & $-{1\over2}$ & ~~0 & $-1$
\|(1,1)\| ~~${8\over3}$ & ~~0 \cr
$D_3$ \|\ $1+\alpha+2\gamma$
\|(1,1,3)\|
{}~~${2\over3}$ & ~~0 & ~~0 & ~~0 & ~~0 & $-1$ & $-1$
\|(1,1)\| $-{4\over3}$ & ~~0 \nr
${\bar D}_3$ \|\
\|(1,1,${\bar3}$)\|
$-{2\over{3}}$ & ~~0 & ~~0 & ~~0 & ~~0 & ~~1 & ~~1
\|(1,1)\| ~~${4\over3}$ & ~~0
\endtable}
\hfill}
\smallskip
\parindent=0pt
\hangindent=39pt\hangafter=1
\baselineskip=12pt
{\it Table 3.}
Massless states and their quantum numbers in the model of table 1.

\vfill
\eject

\bye

\end